\documentclass[preprint,3p]{elsarticle}
\usepackage{amssymb}
\usepackage{multicol}
\usepackage{graphicx}
\usepackage{epsfig}
\usepackage{fancybox}
\usepackage{pstricks,pst-node,pst-text,pst-3d}
\usepackage{subfigure}
\begin{document}
\begin{frontmatter}
\title{Hysteretic response characteristics and dynamic phase transition via site dilution in the kinetic Ising model}
\author[deu]{B.O.~Akta\c{s}\fnref{fn1}}
\author[deu]{{\" U}.~Ak\i nc\i}
\author[deu]{H.~Polat\corref{cor1}}

\cortext[cor1]{Corresponding author: Tel:+90-232 3018672 Fax:+90-232 4534188}
\ead{hamza.polat@deu.edu.tr}
\address[deu]{Department of Physics, Dokuz Eyl\"{u}l University, TR-35160 Izmir, Turkey}
\fntext[fn1]{Also at Dokuz Eyl\"{u}l University, Graduate School of Natural and Applied Sciences, TR-35160 Izmir, Turkey.}


\begin{abstract}
The decay of the hysteresis loop area of the system, which is obeying a site diluted kinetic Ising model, is considered by the disorder parameter using the effective field theory analysis. The exhibition focuses on the understanding of external field frequency, amplitude and the site concentration dependency of the hysteresis loop area for several powerful treatments. Important characteristics of the hysteretic response, such as frequency dispersion, effect of domain nucleation phenomenon on the dynamic process etc. has been introduced together with well known other characteristics. An attempt has been made to explain the relations between the competing time scales (intrinsic microscopic relaxation time of the system and the time period of the external oscillatory field) and the shape of the response. As a result of the detailed investigations, existence of essentially three, particularly four types of dispersion curves have been propounded.
\end{abstract}
\end{frontmatter}

\section{Introduction}\label{introduction}
When a cooperative many-body system such as a magnet is subject to an alternating external magnetic field, the ordinary magnetization, in other words, time dependent magnetization may not respond to the external field simultaneously. Hence the system exhibits an ubiquitous phenomenon which is called hysteresis. The phenomenon is a signature of how the system parameters respond dynamically to the external field sweep. For the kinetic Ising-type systems, the most familiar hysteresis in the form of Lissajous curve is $m(t)$ of the system versus the external magnetic field $h(t)$. Early experimental studies involving ground-breaking discoveries in the understanding of hysteresis on real magnetic systems by Steinmetz, dating back to last century, there is an empirical law for the hysteresis loops (HLs) areal scaling \cite{steinmetz}. Also, the hysteresis loop area (HLA) has been analyzed as a function of the two open variables of the external magnetic field (amplitude and frequency) by fitting the data from large-scale (wide-range) to an areal Lorentzian and/or different scaling law with from low values of the relevant variables to a monotonically decreasing law proposed at various times in outstanding papers in the past two decades by different researchers \cite{rao1, jung, luse, acharyya1, acharyya2, acharyya3, fan, chakrabarti, liu1, liu2}. Recently, many efforts have been devoted to prediction and experimental verification of the scaling behavior of the HLA for thin magnetic films (for a brief review of HLA scaling results see \cite{suen}). Apart from these, there are detailed geometrical description \cite{mayergoyz}, a statistical theory of the nonlinear relaxation function to describe metastable decay \cite{binder}, and hysteresis criterion based on rate competition \cite{gilmore, agarwal}. Additionally, the theoretical observations in the relevant references \cite{jung, luse, acharyya4} show that the HLs has a nonzero residual loop area in the zero-frequency limit (quasistatic limit). The relevant unassailable experimental observation \cite{bruno} is also in agreement with the existence of such residuality. Hysteretic behavior is, however, a complicated process of a dynamic and nonlinear nature that elude serious treatment, both experimentally and theoretically since the area enclosed by the HL is directly proportional to the energy loss in a magnetization-demagnetization cycle.

HL originates from a competition between time scales of relaxation treatment of the system and the periodic external magnetic field. At high temperatures for the high amplitudes of the external magnetic field the system follows the external field with a delay which is called phase lag while this is not the case for relatively low temperatures and small amplitudes. This spontaneous symmetry loss indicates the presence of a dynamical phase transition (DPT) which shows itself in dynamical order parameter (DOP) which is defined as the time average of the magnetization over a full period of the oscillatory field \cite{tome}. DPT is firstly observed theoretically by using the master equation associated to the Glauber-type stochastic process \cite{glauber} within the framework of mean field approximation (MFA), and it follows that the time dependent magnetization satisfies the Suzuki-Kubo mean field equation \cite{suzuki}. Tom\`{e} and Oliveira \cite{tome} located a dynamic tricritical point (DTCP) where the type of DPT changes across the dynamical phase boundary in their pioneering work. However, the first-order dynamical transitions are accepted as artifact by some researchers in the subsequent years \cite{korniss, berkoilako, shi1, idigoras}. Since then, several experimental studies have been put forward to explain the dynamic nature of the phase transition and its mechanism. As an example, a DPT occurring in high frequency region is represented by Jiang et al. \cite{jiang} using the surface magneto-optic Keer-effect technique for epitaxially grown ultrathin Co films on a Cu(001) surface. The same technique has been used by Nowak et al. \cite{nowak} to explain the temperature and field dependence of the magnetization reversal process on Co$_{28}$Pt$_{72}$-alloy samples. In addition, as an experimental example of hysteresis mechanism,  frequency dispersion of the HLA for the ferroelectric material Pb(Zr$_{0.52}$Ti$_{0.48}$)O$_{3}$ has been examined by Liu et al. \cite{liu2}. For a [Co(4A$^{\circ}$)/Pt(7A$^{\circ}$)] multilayered system with a strong perpendicular anisotropy, DPT is shown by Robb et al. \cite{robb}. In the relevant study, they reported that the experimental nonequilibrium phase diagrams are found to strongly resemble the dynamic behavior calculated from simulations of the kinetic Ising model. Hence, the strong evidence of consistency between theoretical and experimental studies can be seen in. Polyethylene naphthalene nanocomposites have been investigated by Kanuga et al. \cite{kanuga}. After detailed analysis, they showed that the material undergoes three critical structural transitions. Up to now from the past periods, it can be seen that increasing of the activities focusing on the most current statement on DPT by researchers using Monte Carlo (MC) simulations \cite{rao1, acharyya3, acharyya4, lo, acharyya5, sides, zhu}, Effective Field Theory (EFT) \cite{shi2, deviren} and MFA \cite{acharyya3, acharyya4, zhu, punya}. In some of them, several methods were used as comparison. By employing the standard Metropolis MC algorithm with periodic boundary conditions, the kinetic pure Ising model in a two dimensional square lattice has been simulated by Lo and Pelcovits \cite{lo}, and they found evidence for a DPT. After a comprehensive investigation, firstly in the literature showed by Acharyya and Chakrabarti that the HLs are asymmetric in the dynamically ordered phases while they are symmetric in the disordered phases around the origin \cite{acharyya3}. Subsequently, Acharyya \cite{acharyya4} presented that the dynamic coercive field exhibits a power law frequency variation both in the MFA and MC simulation after high precision numerical and computational effort. However, the coercive field calculated by using MFA becomes frequency independent in the quasistatic limit. This phenomenon is explained by the author constructing the Landau-type double-well free energy mechanism with the thermal fluctuations and nucleation process and specified that the coercivity can be thought as a limit of metastability. The similar construction has been done between double-peak frequency dispersion with the newly defined two parameters which correspond to the domain-nucleation and -growth rates respectively by Zhu et al. \cite{zhu}. In the first one of the two studies published in the recent years, Shi et al. \cite{shi2} concluded that the system does not contain the coexistence region where dynamically ordered and disordered phases coexist. Following the same methodology, a detailed investigation of the kinetic Ising system has been done by Deviren et al. in the second one \cite{deviren}. They found that the system exhibits the coexistence region which strongly depends on the frequency and amplitude of the external magnetic field. In the same year, Punya et al. \cite{punya} represented the phase diagram in ($k_{B}T_{c}/J-h_{0}/J$) plane for varying frequencies, in order to show the influences of the frequency on the system and they presented in depth satisfactory knowledge about the sweeping in time. Additionally, they reported the frequency dispersion of the HLA, the remanence and coercivity can be categorized into three distinct types for a fixed temperature. However, the article contains no comments about the domain nucleation process. The analytically categorization of the HLs propounded by Lyuksyutov et al. \cite{lyuksyutov}. In addition to this, the two other works about the relations between the domain wall motion and the dynamic hysteresis, are published by approximately the same collaborators \cite{nattermann1, nattermann2}. In the current year, Idigoras et al. \cite{idigoras} have demonstrated that the bias field behaves as if it is the conjugate field of the DOP near the DTP for ferromagnets. Also, they showed that DOP which has the conjugate bias field, follows a power law behavior at critical frequency and its critical exponent is in agreement with the equilibrium MFA value. Despite the fact that the thermal and magnetic properties of the quenched disorder systems such as site (or bond) diluted ferromagnets driven by a periodically oscillating magnetic field have been examined within the framework of EFT very recently in two works \cite{akinci, vatansever} but then as far as we know, the maximum lossy behavior of the such systems have not yet been investigated. For the site diluted kinetic Ising model, the global phase diagrams including the reentrant phase transition are presented by the authors. They showed that the system exhibits a DTCP and coexistence region both of which disappear for sufficiently weak dilution \cite{akinci}. Following the same methodology, the relevant collaborators revealed that the impurities in the system has a tendency to destruct the first-order transitions and the DTCP for the bond diluted kinetic Ising model \cite{vatansever}. And also, they showed that dynamically ordered phase regions get expanded with decreasing amplitude which is more evident at low frequencies.

It is well known that the EFT is one of the most powerful method that determines the meniscus which separates several phases in the relevant planes, based on the use of rigorous correlation identities as a starting point and utilizes the differential operator technique firstly developed by Honmura and Kaneyoshi \cite{honmura}. Although the conventional version of the method fails to find an expression for the free energy, due to it takes into account the self spin correlations, the method is superior to MFA which neglects the thermal fluctuations via neglecting the self spin correlations. Thus, it is expected from EFT to obtain more reasonable results than MFA for these systems, as in the case of static Ising model. Therefore in this work, we intend to probe the effects of the quenched site dilution process on the dynamic hysteresis characteristics of kinetic Ising model in the presence of a time-dependent external oscillatory magnetic field by using the EFT. These types of disorder effects constitute an important role in material science, since the quenched disorder effects may induce some important macroscopic effect which is still open for inspection. Eventually, for these purposes the outline of the article as follows: We briefly describe the formalism and the method used for the problem in Sec. 2. Numerical results and discussions are summarized in Sec. 3, and finally Sec. 4 contains remarks about our conclusions.

\section{Methodology}\label{formulation}
To investigate the dynamical transitions, dynamical symmetry loss (similar to the phrase symmetry breaking in the static Ising model), hysteretic response and many other dynamical features, one simple choice may be an Ising ferromagnet defined on the two dimensional space which has a coordination number $z=3$ with a time dependent external oscillatory (in time but uniform over the space) magnetic field studied by EFT. For this purpose, we consider the following Hamiltonian,
\begin{equation}\label{eq1}
\mathcal{H}=-J\sum_{<ij>}c_{i}c_{j}s_{i}s_{j}-h(t)\sum_{i}c_{i}s_{i},
\end{equation}
where $J>0$ favors a ferromagnetic exchange interaction of the adjacent sites, $c_{i}$ is a site occupation variable and $s_{i}$ is the spin variable. Site occupation variable can take the values $c_{i}=0$ which means that the site $i$ is empty or has a non-magnetic lattice item, $c_{i}=1$ if the site $i$ is occupied by a magnetic item and the spin variable can take values $s_{i}=\pm 1$. The first summation in Eq. (\ref{eq1}) is over the nearest neighbor site pairs and the second one is over all lattice sites. The Zeeman term describes interaction of the spins with the field of the sinusoidal form
\begin{equation}\label{eq2}
h(t)=h_{0}\cos(\omega t),
\end{equation}
where $t$ is the time and $h_{0}$ is the amplitude of the oscillatory magnetic field with a frequency $\omega$. The system is in contact with an isothermal heat bath at given temperature $T$. So, the dynamical evolution of the system may be given by non-equilibrium Glauber dynamics \cite{glauber} based on a master equation
\begin{equation}\label{eq3}
\tau \frac{d}{dt}\langle\langle c_{i}s_{i}\rangle\rangle_{r}=-\langle\langle c_{i} s_{i}\rangle\rangle_{r}+\langle\langle c_{i}\tanh[\beta c_{i}(E_{i}+h(t))]\rangle\rangle_{r},
\end{equation}
where $1/\tau$ is the transition per unit time in a Glauber type stochastic process, $\beta=1/k_{B}T$ and $k_{B}$ represents the Boltzmann constant and $E_{i}$ is the local field acting on the site $i$ and it is given by
\begin{equation}\label{eq4}
E_{i}=J\sum_{\delta=1}^{z}c_{\delta}s_{\delta}.
\end{equation}
The inner average brackets in Eq. (\ref{eq3}) stands for the usual canonical thermal average and the outer one (which has an index $r$) represents the random configurational average which is necessary for including the site dilution effects.

In order to handle the second term on the right hand side of the Eq. (\ref{eq3}) one can use the differential operator technique \cite{honmura, kaneyoshi1}. By using the technique, Eq. (\ref{eq3}) gets in the form of
\begin{equation}\label{eq5}
\tau \frac{d}{d t}\langle\langle c_{i}s_{i}\rangle\rangle_{r}=-\langle\langle c_{i}s_{i}\rangle\rangle_{r}+\langle\langle c_{i}\exp (c_{i} E_{i}\nabla)\rangle\rangle_{r}f(x)|_{x=0}
\end{equation}
where $\nabla=\partial/\partial x$ is one dimensional differential operator and the function $f(x)$ is given by
\begin{equation}\label{eq6}
f(x)=\tanh[\beta(x+h(t))],
\end{equation}
as for that the effect of the differential operator on a function $f(x)$,
\begin{equation}\label{eq7}
\exp(a\nabla)f(x)|_{x=0}=f(x+a)|_{x=0},
\end{equation}
with any real constant $a$. By taking into account the two possible values of $c_{i}$ as $c_{i}=0,1$ we can write the exponential term as
\begin{equation}\label{eq8}
\exp(a c_{i})=c_{i}\exp(a)+1-c_{i}.
\end{equation}
By using the expansion Eq. (\ref{eq8}) in Eq. (\ref{eq5}) we can obtain
\begin{equation}\label{eq9}
\tau\frac{d}{dt}\langle\langle c_{i}s_{i}\rangle\rangle_{r}=-\langle\langle c_{i}s_{i}\rangle\rangle_{r}+\langle\langle c_{i}\exp(E_{i}\nabla)\rangle\rangle_{r}f(x)|_{x=0},
\end{equation}
where the site occupations $c_{i}^{2}=c_{i}$ was used. By inserting Eq. (\ref{eq4}) into Eq. (\ref{eq9}) we get
\begin{equation}\label{eq10}
\tau\frac{d}{dt}\langle\langle c_{i}s_{i}\rangle\rangle_{r}=-\langle\langle c_{i}s_{i}\rangle\rangle_{r}+\left\langle\left\langle c_{i}\prod_{\delta=1}^{z}[c_{\delta}\exp(Js_{\delta}\nabla)+1-c_{\delta}]\right\rangle\right\rangle_{r}f(x)|_{x=0}.
\end{equation}
In order to get a polynomial form of the second term on the right hand-side of the equation, by using the van der Waerden identity for two-state spin, i.e., $\exp(bs_{i})=\cosh(b)+s_{i}\sinh(b)$ where $b$ is any real constant in a representative manner, we write the exponential term in terms of the hyperbolic trigonometric functions, Eq. (\ref{eq10}) exactly written in terms of multiple spin correlation functions occurring on the right-hand side. Thus we get
\begin{equation}\label{eq11}
\tau\frac{d}{dt}\langle\langle c_{i}s_{i}\rangle\rangle_{r}=-\langle\langle c_{i}s_{i}\rangle\rangle_{r}+\left\langle\left\langle c_{i}\prod_{\delta=1}^{z}[c_{\delta}\cosh(J\nabla)+c_{\delta}s_{\delta}\sinh(J\nabla)+1-c_{\delta}]\right\rangle\right\rangle_{r}f(x)|_{x=0}.
\end{equation}
When the product in Eq. (\ref{eq11}) is expanded, the multi site spin correlations appear. In order to make the expansion manageable let us handle these correlations with an improved decoupling approximation (DA) \cite{tucker} as
\begin{equation}\label{eq12}
\langle\langle c_{i}\ldots c_{j}c_{k}s_{k}\ldots c_{l}s_{l}\rangle\rangle_{r}=\langle\langle c_{i}\rangle\rangle_{r}\ldots \langle\langle c_{j}\rangle\rangle_{r}\langle\langle c_{k}s_{k}\rangle\rangle_{r}\ldots \langle\langle c_{l}s_{l}\rangle\rangle_{r},
\end{equation}
with
\begin{eqnarray}\label{eq13}
\langle c_{i}\rangle_{r}&=&c\nonumber\\
\langle\langle c_{i}s_{i}\rangle\rangle_{r} &=&m.
\end{eqnarray}
In fact, the primitive form of this improvement corresponds essentially to the Zernike approximation \cite{zernike} in the bulk problem, and has been successfully applied to a great number of magnetic systems including the surface problems \cite{kaneyoshi1, balcerzak, kaneyoshi2, kaneyoshi3}. Detailed discussion of DA in site dilution problem for the static case can be found in \cite{tucker}. By using the approximation in Eq. (\ref{eq12}) and the definitions in Eq. (\ref{eq13}) in Eq. (\ref{eq11}) we get
\begin{equation}\label{eq14}
\tau\dot{m}=-m+c\langle\langle[c\cosh(J\nabla)+m\sinh(J\nabla)+1-c]^{z}\rangle\rangle_{r}f(x)|_{x=0}.
\end{equation}
By using the binomial expansion and writing the hyperbolic trigonometric functions in terms of the exponential functions we get the most compact form of Eq. (\ref{eq14}) as
\begin{equation}\label{eq15}
\dot{m}=\frac{1}{\tau}\left(-m+\sum_{\nu=0}^{z}\Lambda_{\nu}m_{\nu}\right),
\end{equation}
where
\begin{equation}\label{eq16}
\Lambda_{\nu}=\sum_{p=\nu}^{z}{{z}\choose{p}}{{p}\choose{r}}c^{p-\nu+1}(1-c)^{z-p}\cosh^{p-\nu}(J\nabla)\sinh^{r}(J\nabla)f(x)|_{x=0}.
\end{equation}
Eq. (\ref{eq15}) is typical first order ODE but has not an analytical solution because of the right hand-side contains transcendental functions. The dynamical equation of motion can be solved by various numerical methods. In this work, we prefer to use the fourth order Runge-Kutta method (RK4) to get the evolution of the $m(t)$ by regarding Eq. (\ref{eq15}) as an initial value problem. We can mention that the differential equation derived in Eq. (\ref{eq15}) extends up to the term $m^{z}$. Each term in the equation of motion makes contribution to the solution, because the value of $m$, which is calculated at each time step, is iteratively related to the previous $m$ value, however the situation is different from the behavior of the equilibrium systems at which high ordered terms can be neglected in the neighborhood of phase transition point.

The system has three dependent Hamiltonian variables, namely frequency of external magnetic field ($\omega$), amplitude ($h_{0}$) and the dilution parameter ($c$). For certain values of these parameters and temperature, RK4 will give convergency behavior after some iterations i.e. the solutions have property $m(t)=m(t+2\pi/\omega)$ for arbitrary initial value for the magnetization ($m(t=0)$). Each iteration, i.e. the calculation magnetization for $i+1$ from the the previous magnetization for $i$, is now performed for these purpose whereby the RK4 iterative equation is being utilized to determine the magnetization for every $i$. In order to keep the iteration procedure stable in our simulations, we have chosen $10^{4}$ points for each RK4 step. Thus, after obtaining the convergent region and some transient steps (which depends on Hamiltonian parameters and the temperature) the DOP can be calculated from
\begin{equation}\label{eq17}
Q=\frac{\omega}{2\pi}\oint m(t)dt
\end{equation}
where $m$ is a stable and periodic function anymore. As cut-off condition for numerical self-consistency, we defined a tolerance
\begin{equation}\label{eq18}
\left|Q|_{-2\pi/\omega}^{t}-Q|_{t}^{2\pi/\omega}\right|<10^{-6}
\end{equation}
meaning that the maximum error as difference between the each consecutive iteration should be lower than $10^{-6}$ for all step. On the other hand, the HLA corresponding to energy loss due to the hysteresis is defined as,
\begin{equation}\label{eq19}
A=-\oint m(t)dh(t)=h_{0}\omega \oint m(t)\sin(\omega t)dt.
\end{equation}

There are three possible states for the system, namely ferromagnetic (F), paramagnetic (P) and the coexistence phase (F+P). The solution $m(t)$ in convergent region is satisfied by the tolerance
\begin{equation}\label{eq20}
m(t)=-m(t+\pi/\omega)
\end{equation}
in the P phase which is called the symmetric solution. The solution corresponding to P phase follows the external magnetic field and oscillates around zero value which means that the DOP is zero. In the F phase, the solution does not satisfy Eq. (\ref{eq20}) and this solution is called as non-symmetric solution which oscillates around a non-zero magnetization value, and does not follow the the external magnetic field i.e. the value of $Q$ is different from zero. In these two cases, the observed behavior of magnetization is regardless of the choice of initial value of magnetization $m(0)$ whereas the last phase has magnetization solutions symmetric or non-symmetric depending on the choice of the initial value of magnetization corresponding to the coexistence region where F and P phases overlap. The remanent magnetization (in other words residual magnetization which is the magnetization left behind in the system after an external magnetic field is removed) and coercive field which means that the intensity of the external magnetic field needed to change the sign of the magnetization, can be calculated by benefiting from the HL in order to understand and clarify the behavior of the site diluted dynamical system.

\section{Results and Discussion}\label{results}

\begin{figure}
\center
\includegraphics[width=14cm]{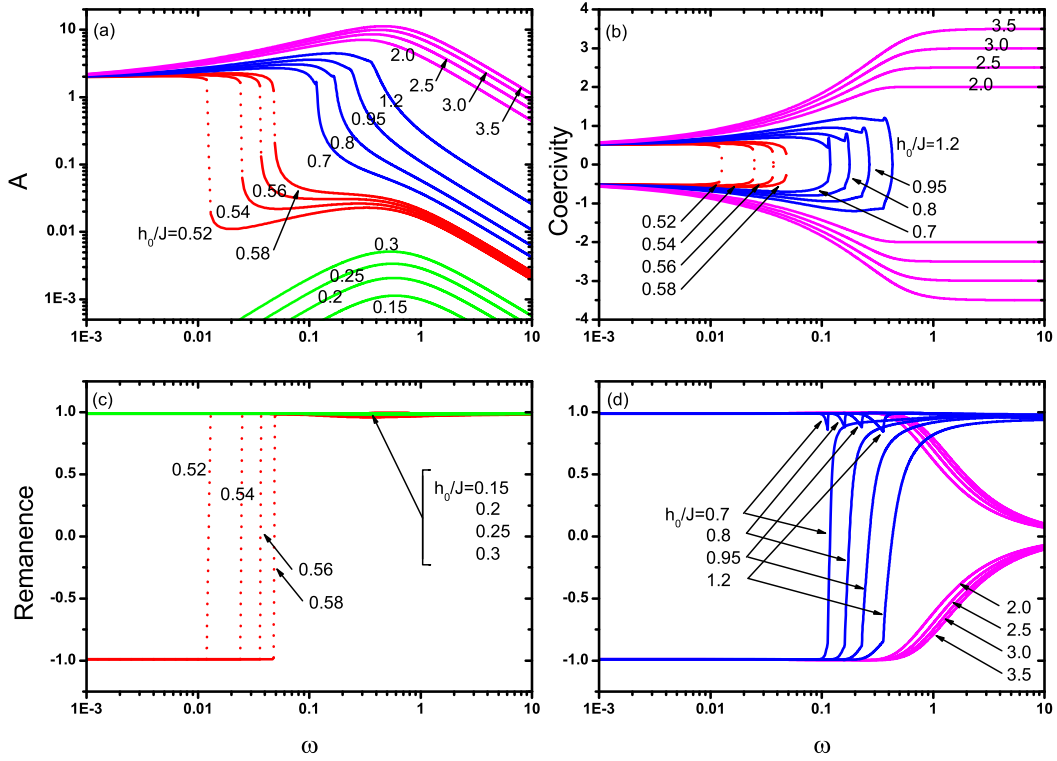}\\
\caption{The frequency dispersion of (a) HLA, (b) coercive field and (c)(d) the remanent magnetization for various $h_{0}/J$ at $T=0.5T_{c}$, which are classified into four types of formal care: Type I at $h_{0}/J=0.15, 0.2, 0.25, 0.3$, II.a at $h_{0}/J=0.52, 0.54, 0.56, 0.58$, II.b at $h_{0}/J=0.7, 0.8, 0.95, 1.2$ and type III at $h_{0}/J=2.0, 2.5, 3.0, 3.5$ respectively. The lines serve as a viewing guide.}\label{fig1}
\end{figure}

The frequency dispersion, the corresponding coercivity and remanent magnetization for various $h_{0}/J$ at $T=0.5T_{c}$ ($0.5T_{c}\cong 1.045$ where $T_{c}$ is the Curie temperature of EFT DA solution with coordination number $z=3$ in static case) for a pure ferromagnet which has been rigorously depicted in the text, constitutes as the starting point for our systematic investigation, are presented in Fig. (\ref{fig1}). With the guidance of the brief report presented by Zhu et al. \cite{zhu}, as a result of evaluations by the conviction that we have reached, the HLA frequency dispersion can be classified into (in particular) four types as type I, II.a, II.b and III, and plotted with different color in each panel. In the panel (a) of Fig (\ref{fig1}), type I has single peak functions which is labeled as peak II (PII) from the asymmetric HLs, II.a where the area suddenly changes from a symmetric to an asymmetric HLs and there is pseudo-double peak functions, PI (why it is not taken into consideration for this type of curves will be described below) and the PII (which is prominently in the asymmetric region), II.b where there are markedly double peak functions, PI and PII which are corresponding to smooth directional-veering frequency point and symmetry loss of the HLs respectively. In the type II.b, as the frequency is increased from PI to PII, a dynamic symmetry loss occurs continuously. And finally, we see that the type III has a single peak functions PI from the symmetric HLs with a finite value (residual area) in quasistatic limit. We also point out that this labelling procedure can be used for all relevant behaviors in different parameter dependency, we have also made it for the concentration dependency. Low coercivity is desired in the materials, which means low hysteresis loss per cycle of operation. Based on this fact, we have examined both coercivity and remanent magnetization. In the panel (b) of Fig. (\ref{fig1}), type II.a coercivity appears only at low frequencies since at a critical frequency ($\omega_{c}$) values, the hysteresis loop is asymmetric and does not cross the fixed-axis. In the II.b of the types, again the coercivity appears relatively at low frequencies by a difference from type II.a characteristics: The symmetry loss phenomenon takes place in the PII and the coercive field has maximal magnitute at that frequency value. In principle, from the relevant veering till the symmetry loss the dynamics controlled by so-called reversal field and at the end of this process a gibbous-like eminentia appears as seen in the right arm of the mutual coercivities. Type III coercivity is small at low frequencies but gets larger and approaches the amplitude at higher frequencies. The type I coercivity are not shown, because no coercive field is found due to the asymmetric behavior. In the panel (c) for the remanent magnetization which is belonging to the type I collection, we see that there exists only an asymmetric loop. For the type II.a and II.b in the panel (c) and (d) respectively, both positive and negative remanence appears at low frequencies, but only positive values appear at high frequencies due to the symmetry loss. However, they are separated by a difference such as symmetry loss is discontinuous or continuous. Type III in the panel (d), both positive and negative remanence values appear at all the shown frequencies.

\begin{figure}
\center
\includegraphics[width=8cm]{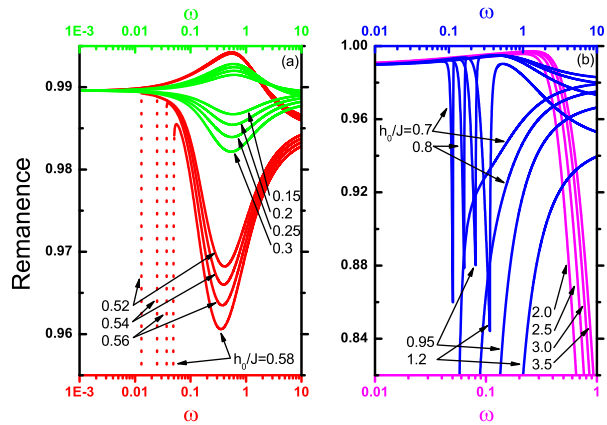}\\
\caption{(a) Type I, II.a and (b) II.b, III remanences at $T=0.5T_{c}$ zoom in as plotted again in terms of visual clarity. The numbers accompanying each curve denote the value of the amplitude of the external oscillatory field.}\label{fig2}
\end{figure}

Shown in Fig. (\ref{fig2}) plotted for better visibility, why the type II frequency dispersion curves which are categorized by Punya et al. \cite{punya}, is classified into two labeled as II.a and II.b as an extra. As can be seen, the most remarkable types of hysteresis related to frequency dispersion are these two particular types, which contains features from both types I and III. It clearly shows the effect of frequency on the dynamic HLA and dynamic phase. Unlike other stochastic methods such as MC simulation technique which takes into account thermal fluctuations, there is a very important difference in EFT DA solution of the remanent. Given relatively small values of $h_{0}$, it is possible that there is only PII in the EFT solution. This is because of that the equilibrium can be reached by solving the dynamical equation of motion with $h_{0}$ small enough and $T<T_{c}$, there can be two real stable time series solutions which oscillate around a non-zero value, corresponding to two values of stable equilibrium DOP (a positive and negative one). This means that the HLs can be asymmetric even in the quasistatic limit. Just a single peak has been encountered along the frequency scale, because all the type I curves are members of the purely asymmetric collection. Simply seen that the steep jump of the type II.a remanence before the local maxima due to PII. Because of the value of mutual remanences almost unchanged up to the symmetry loss, the peak I is not taken into consideration. In the type II.b characteristics, while the positive remanence temporally kinks into the deep, negative remanence gets into the window slightly at that small frequency region. This phenomenon means that continuous symmetry loss exists due to the PII of the HLA dispersion and the eminentia of the coercivity at the same time. Similar to the type I, the type III curves which are member of the purely symmetric collection, have the same characteristics and behaves in the same way. For a fixed parameter set, the relative maximal loop area is observed in this type of frequency dispersion. Also, in the first two figures, the $h_{0}$ dependencies of the HL, coercivity and remanence are considered. If we look at both Fig. (\ref{fig1}) and Fig. (\ref{fig2}) as a whole, at first sight we can see that as $\omega$ value increases then the symmetry loss phenomenon occurs for the specific $h_{0}$ values. This is an expected result, since the increase in frequencies enhances the phase lag between time dependent magnetization and the external field signals. Therefore, the asymmetric behavior of the HL becomes more prominent, since the time dependent magnetization has less time to follow the oscillatory field. Hence, the system can undergo a DPT which requires a small amount of thermal energy. This mechanism is due to the competing time scales in such equilibrium systems and also can be seen in the global phase diagrams in ($k_{B}T_{c}/J-h_{0}/J$) planes presented by Akinci et al. \cite{akinci}. There is an exact consensus in literature that the HL loses its symmetry when the time period of the oscillating external perturbation becomes much smaller than the typical relaxation time of the system. Notwithstanding that an increasing in $h_{0}$ for a particular $\omega$ causes a decrease in sufficient critical temperature $T_{c}$ value for DPT along the relevant ($k_{B}T_{c}/J-h_{0}/J$) transition curve, the same operation for a particular  $T=0.5T_{c}$ causes an increase in sufficient critical $\omega_{c}$ value for the frequency-induced DPT.

\begin{figure}
\center
\includegraphics[width=14cm]{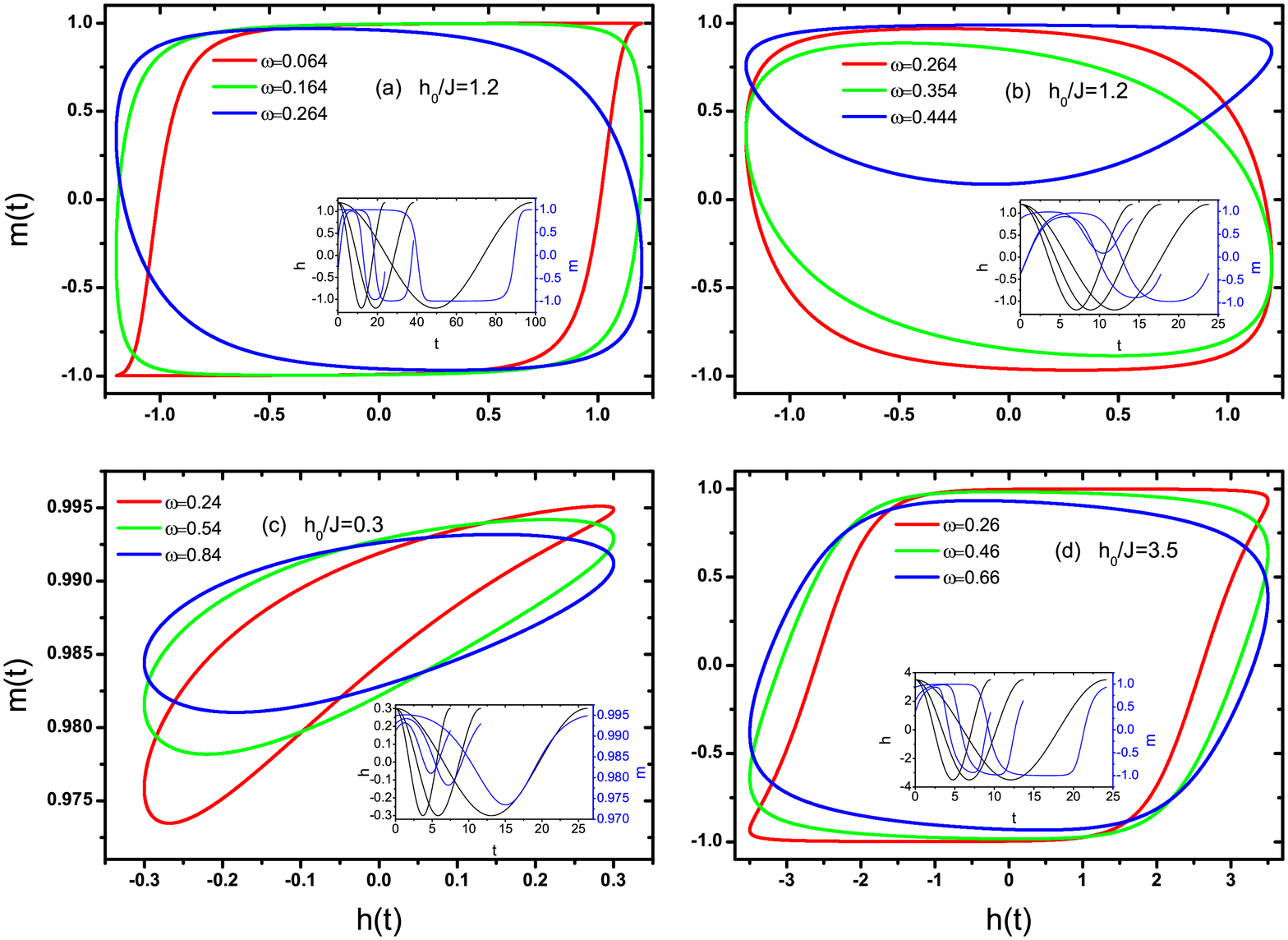}\\
\caption{The HLs in different characteristics for various $h_{0}/J$ at $T=0.5T_{c}$ which are located along the frequency dispersion representatives from (a)(b) type II.b, (c) I and (d) III are shown in the figure. There is no HLs for the representative which is selected from type II.a, because its behavior is exhaustively depicted in the text. Each insets show magnetization component against the instantaneous field component over a time period for the related frequency values.}\label{fig3}
\end{figure}

For a better clarification of the relation between the nucleation process and the frequency-induced DPT, should be consulted to the hysteretic behavior and its frequency evolution in Fig. (\ref{fig3}). Hysteresis loops in fundamentally different characteristics relating to principal three agents for various $h_{0}/J$ at $T=0.5T_{c}$, from the different type of frequency dispersion are shown in the figure. In both panels (a) and (b) shows type II.b at $h_{0}/J=1.2$. Type I at $h_{0}/J=0.3$ and III at $h_{0}/J=3.5$ are seen in the panel (c) and (d) respectively. Additionally, each panel have an inset which shows magnetization component against the instantaneous field component over a time period for selected from the frequency points as experienced significant anomalies. As seen in (a) for $h_{0}/J=1.2$ we observe a smooth directional-veering instead of symmetry loss at $\omega=0.164$. Most importantly, two branches of the hysteresis curve intersect each other at this point and the HLA has the global maxima due to the multiply of its two-axis. Also, the magnetic moment remains an approximately constant value. A little bit the left and the right of this point at $\omega=0.064$ and $0.264$ respectively, corresponding hysteresis curves are topologically contra-oriented mutually and completely different growth mechanism is onset. For the same $h_{0}/J=1.2$ value in (b) as the frequency increases starting from the value at the maximum area of the HL the system undergoes a continuous DPT due to the incomplete reversal of magnetic moment. For the critical frequency values at $\omega>0.354$ all the HLs are in the form of asymmetric. According to times-series of the magnetization and the external magnetic field, increasing the field frequency at first, obstructs the saturation of the ordinary magnetization due to the decreasing energy coming from the oscillating magnetic field in a half-time period which facilitates the late stage domain growth by tending to align the moments in its direction (i.e. the magnetization begins to fail following the oscillatory field) and this makes the occurrence of the frequency increasing route to DPT at the critical point. Purely asymmetric, left-facing teardroplike HLs which belongs to the type I class seen in Fig. (\ref{fig3}) (c). We observed the maximal area at $\omega=0.54$ for $h_{0}/J=0.3$ and there is no DPT due to the increase of the frequency and corresponding relevant time-series of the magnetization which oscillate around a non-zero value as seen in the inset. Finally in panel (d) of Fig. (\ref{fig3}), can be seen the purely symmetric HLs from III and corresponding magnetization oscillation around the non-zero value in the relevant inset. As we can see, a lower frequency range, the magnetization can be more saturated and this shows itself as the dominant magnetization processes in the different regions of the curve (i.e. reversible boundary displacements, irreversible boundary displacements and magnetization rotation). Furthermore, both types I and III show that the HLA increases at a low frequency and then decreases at a higher frequency. The most significant distinction of these two types can be seen in the upper and lower frequency limits behavior, considering the differences between types I and III at $\omega\rightarrow 0 $, for type I the HLA goes to zero, while for type III goes to a finite non-zero value. There are no agents from the type II.a, just because its representative HLs's frequency dependent evolution is similar to the panel (b) by a difference that there is no directional-veering scenario while as the symmetry loss occurs discontinuously.

As can be understood from the results described above, two peaks observed PI and PII which corresponds to the resonance of symmetric and asymmetric HLs with the period of external field, respectively by means of the EFT calculations. According to Gunton and co-workers \cite{gunton}, the dynamics may consist of domain nucleation and/or domain growth such systems that we examined in this study. From this point of view, in the symmetrical region of the HLs, both the initial domain nucleation and the late stage domain growth are onset, hence the system dynamics is controlled by the combined domain nucleation-and-growth process. Esoterically, the type II.a and II.b mechanisms are different from each other. Since for the asymmetric HLs region the DOP can be always well above or below zero, at any time during the system evolution we can observe most magnetic moments having the same direction. The late stage domain growth is relatively inhibited, so the dynamics may be mainly domain nucleation. Lyuksyutov and co-workers explained in detail these process mentioned-above by introducing four characteristic fields \cite{lyuksyutov}. According to their declaration, the process of magnetization reversal can be divided into two stages: Nucleation of domains with opposite magnetization and growth of these domains. Which process dominates depends on system parameters and their history. Our observations are in accordance with these explanations and arguments with some unpretentious improvements.

\begin{figure}
\center
\includegraphics[width=11cm]{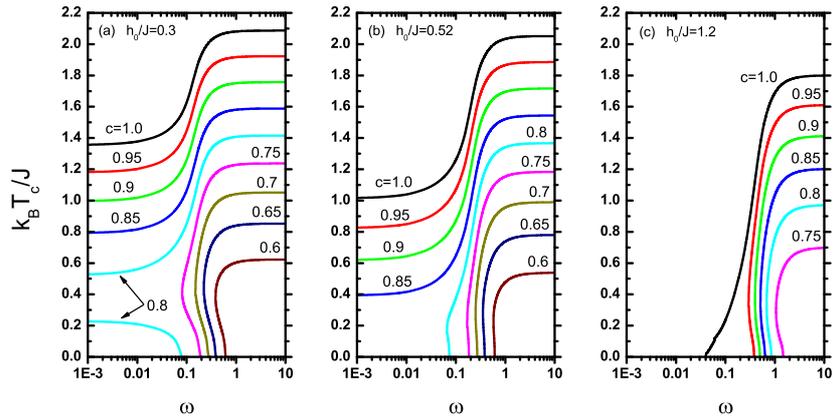}\\
\caption{Phase diagrams for the lattice which has been defined with coordination number $z=3$ in ($k_{B}T_{c}/J-\omega$) plane for the principal three agents stated in the text. The numbers accompanying each represent value of the site concentration.}\label{fig4}
\end{figure}

\begin{figure}
\center
\includegraphics[width=14cm]{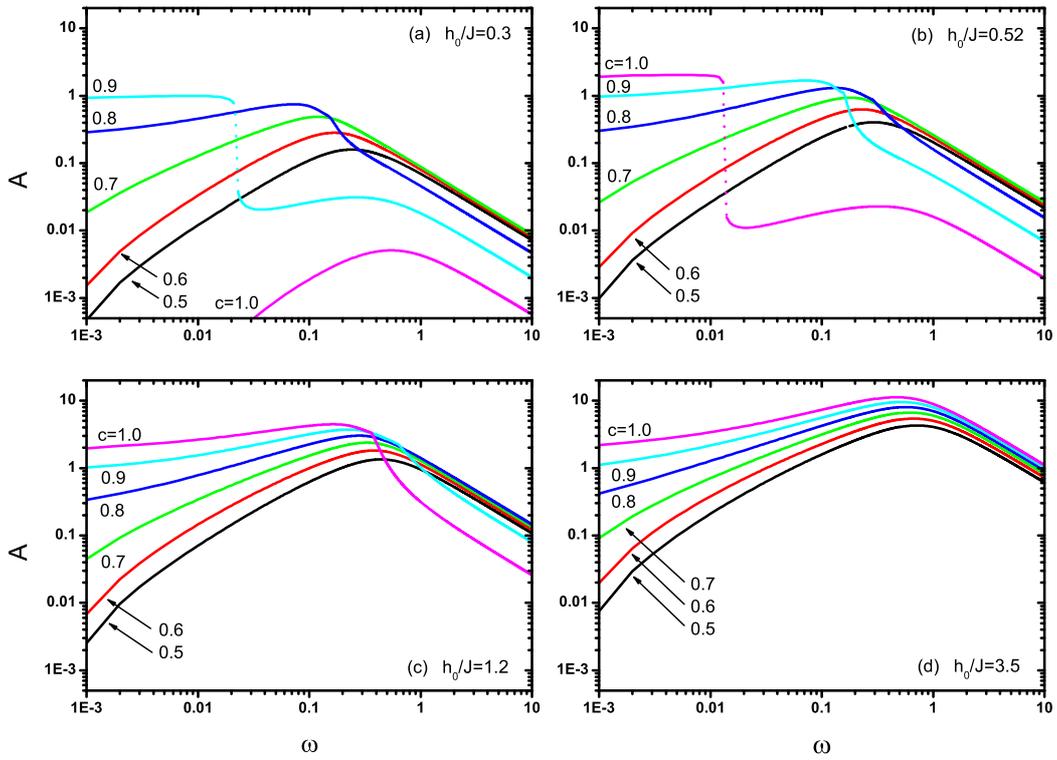}\\
\caption{Disorder-induced evolution of the (a) type I, (b) type II.a, (c) type II.b and (d) type III representatives frequency dispersion at $T\cong 1.045$. The numbers accompanying each represent value of the site concentration.}\label{fig5}
\end{figure}

\begin{figure}
\center
\includegraphics[width=14cm]{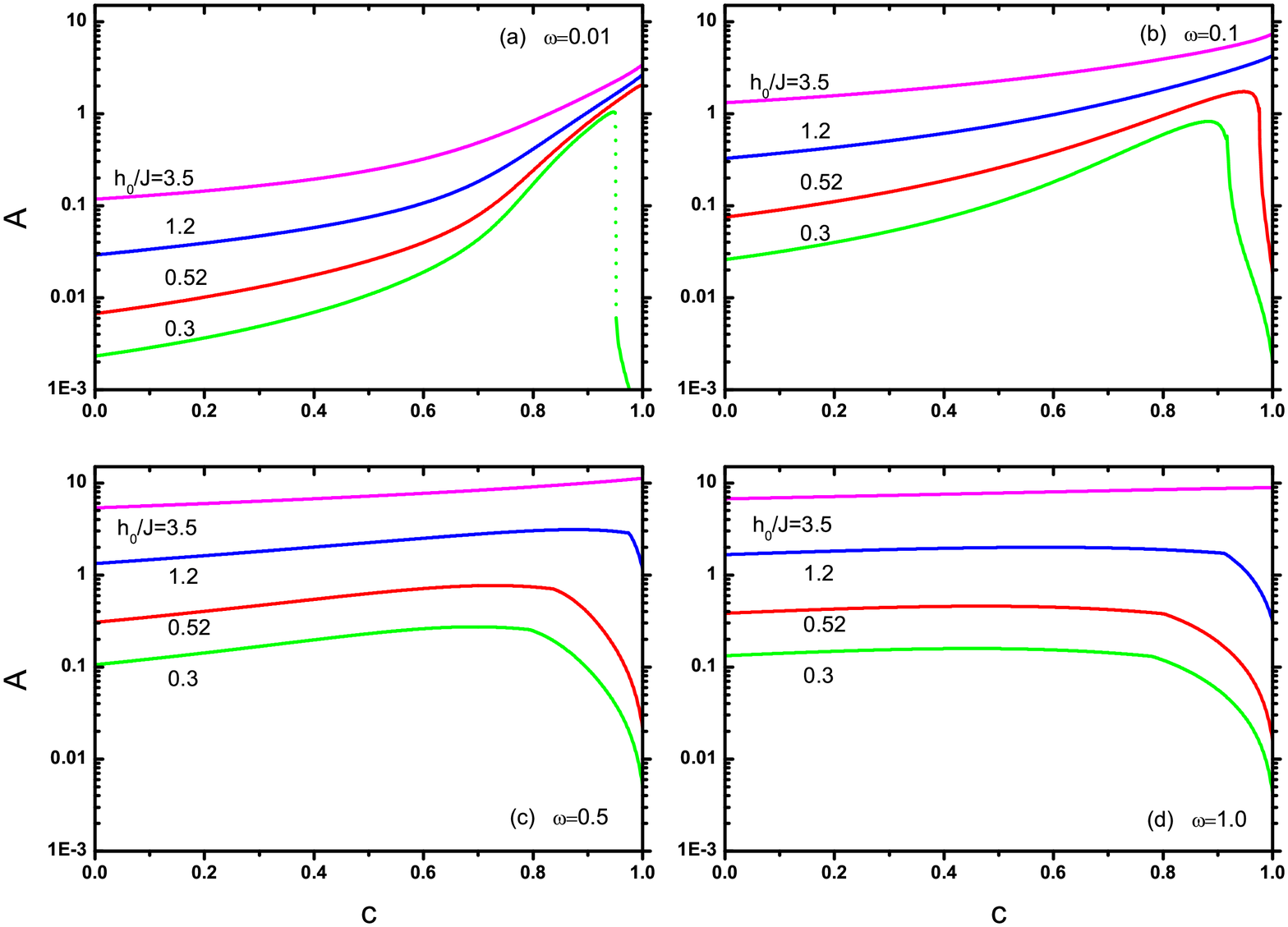}\\
\caption{Variation of the HLA with site concentration $c$ for the selected agents from four different types of the frequency dispersion curves at constant temperature $T\cong 1.045$, for the values of the frequency $\omega=0.01, 0.1, 0.5$ and $1.0$. The numbers accompanying each curve denote the value of the amplitude of the external oscillatory field.}\label{fig6}
\end{figure}

\begin{figure}
\center
\includegraphics[width=14cm]{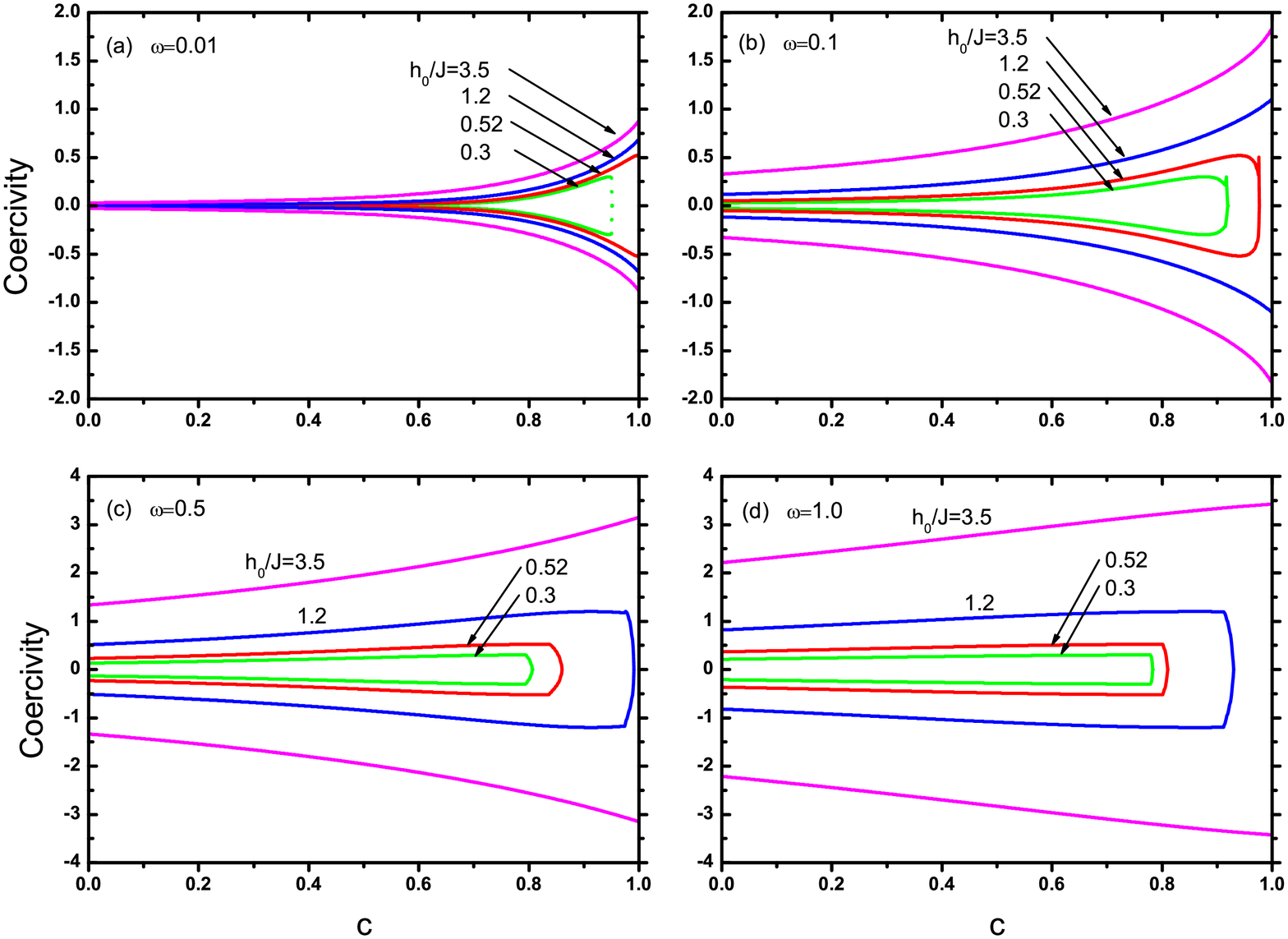}\\
\caption{Variation of the coercivity with site concentration $c$ for the selected agents from four different types of the frequency dispersion curves at constant temperature $T\cong 1.045$, for the values of the frequency $\omega=0.01, 0.1, 0.5$ and $1.0$. The numbers accompanying each curve denote the value of the amplitude of the external oscillatory field.}\label{fig7}
\end{figure}

\begin{figure}
\center
\includegraphics[width=14cm]{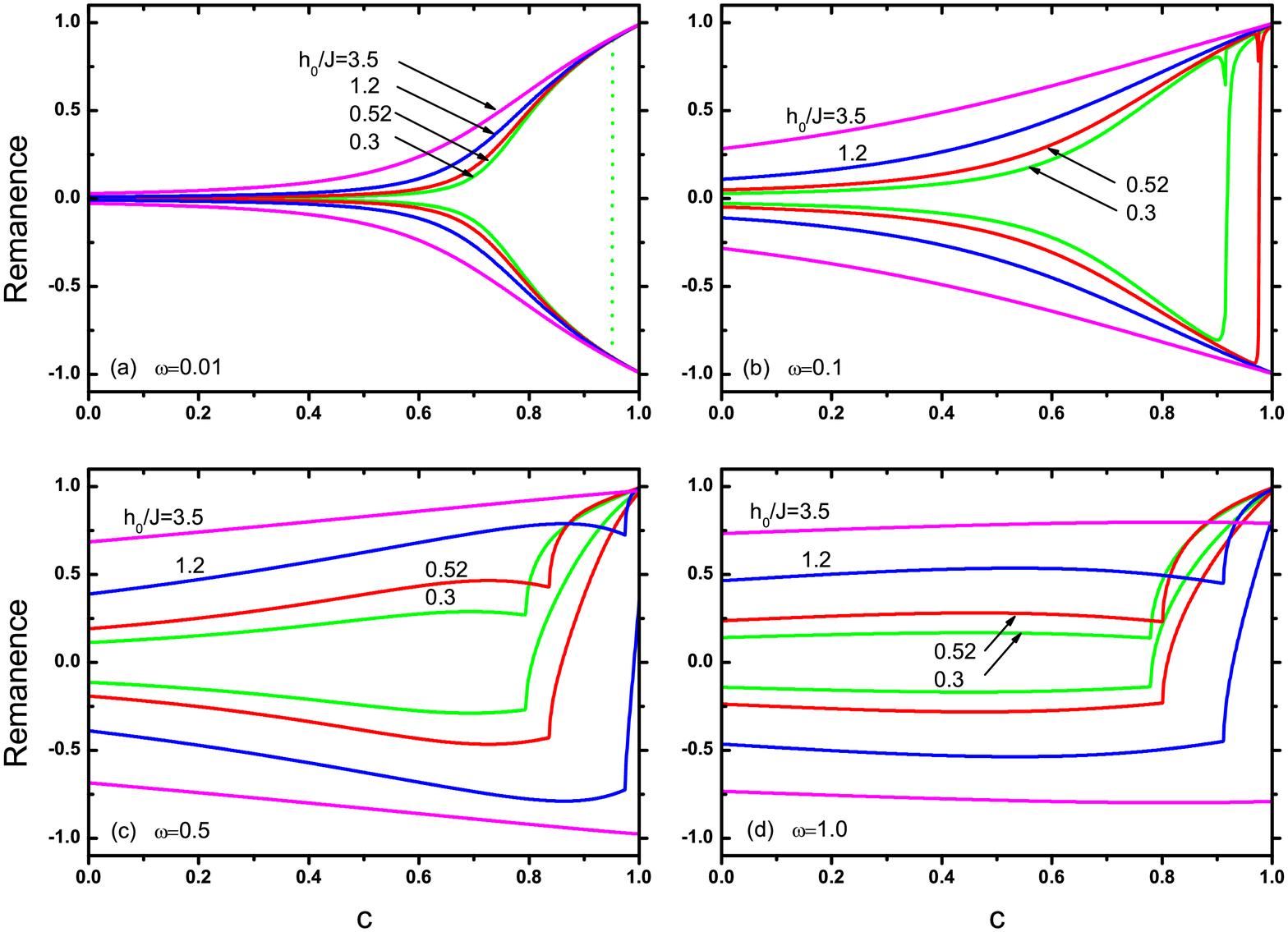}\\
\caption{Variation of the remanence with site concentration $c$ for the selected agents from four different types of the frequency dispersion curves at constant temperature $T\cong 1.045$, for the values of the frequency $\omega=0.01, 0.1, 0.5$ and $1.0$.  The numbers accompanying each curve denote the value of the amplitude of the external oscillatory field.}\label{fig8}
\end{figure}

As one of the most striking aspects of our work is that the dispersion curves evolve from the type I to III hierarchically as site concentration decreases for every $h_{0}/J$ values. As a beginning, we need to dart a glance at the relevant phase diagrams in Fig. (\ref{fig4}) which is plotted for the principal three agents for various values of the amplitude $h_{0}/J$ of the external oscillatory field and the site concentration $c$ so as to examine the evolution and underlying mechanism. The agents which could respond to the site dilution the most quickly are selected. These are (a) type I at $h_{0}/J=0.3$, (b) II.a at $h_{0}/J=0.52$ and (c) II.b at $h_{0}/J=1.2$. The diagram for the type III agent is not shown, because no critical temperature is found due to the asymmetric behavior. A portion of the curves along the selected route at $T\cong 1.045$ are remains the purely asymmetric or remains the purely symmetric region and a small amount of them are signify symmetry loss. In order to focus on this interesting disorder-induced evolution we depicted the frequency dispersions for $h_{0}/J=0.3, 0.52, 1.2, 3.5$ with the concentration values $c=1.0, 0.9, 0.8, 0.7, 0.6, 0.5$ in Fig. (\ref{fig5}). As seen in the figure, at $T\cong 1.045$ for a particular disorder parameter, although, a weak increasing for $h_{0}/J$ causes an increase in $\omega_{c}$, a decreasing of the site concentration causes an increase in $\omega_{c}$ and vice versa. This is an expected results, since as the lattice sites are diluted more and more then the dipole-dipole interaction-induced energy contribution gets smaller. The quantity of the energy contribution which comes from the dipole-dipole interaction corresponds to the self-volition to remain in the dynamically symmetric phase of the system. Hence, the system can stay in the paramagnetic phase in the smaller frequency until the frequency of the external field becomes greater than the relaxation time of the system. In order to probe the concentration dependence of the dynamic nature of the system, we plotted the HLA versus dilution parameter $c$ in Fig. (\ref{fig6}) for various frequency values (a) $\omega=0.01$, (b) $0.1$, (c) $0.5$, and (d) $1.0$ at $T\cong 1.045$. Correspondingly, the unmediated concentration dependency of the coercive field and the remanence magnetization plotted in Fig. (\ref{fig7}) and (\ref{fig8}) respectively for the same parameter set acceptances. In accordance with our anticipation, as the concentration $c$ of magnetic atoms decreases then the system puts under the command of the symmetric behavior. In other words, decreasing of the concentration facilitates the symmetry tendency of the system. As a new characteristic behavior, such as discontinuous symmetry losses are also available in this anthologia for $h_{0}=0.3$ at $\omega=0.01$. We can also observe the existence of the continuous symmetry loss and correspondingly above-cited peak I and II (related eminentia). As a general trend, coercivity is small at low site concentrations, but gets larger and approaches the amplitude at higher concentration. As seen in Fig. (\ref{fig7}), increasing the frequency causes the behavior gradually becomes linear and in Fig. \ref{fig8} both positive and negative remanence appears at low frequencies, but only positive values appear at high frequencies due to the symmetry loss. The discontinuous symmetry loss, also the kinks corresponding to the relevant eminentia etc.  can be clearly seen. Moreover, the low amplitudes concentration dependencies in Fig. (\ref{fig6}), (\ref{fig7}) and (\ref{fig8}), which appear at small $\omega$ values such as $\omega=0.01$, exhibit a discontinuous symmetry loss and the new type II.b curves reproduce hierarchically from II.a curves as increasing frequency. This phenomena can seen as the evolution from the sudden symmetry loss to coercivity which has two peaks in Fig. (\ref{fig7}) and the steep jump behavior to the kink of the remanence in Fig. (\ref{fig8}). All statements concerning the mechanism of this behavior, also apply here. Furthermore, HLA goes to a finite value in the limit of $c\rightarrow 0$. It means that, no matter the system exhibit a long range ordered phase or not, for all concentration values we can observe the HLA, correspondingly coercivity and remanence for concentration values even below its critical value $c^{*}$ (namely site percolation threshold).

Concretely speaking, in order to determine the type of the frequency or disorder-induced symmetry loss (continuous or discontinuous), we appeal to hysteresis for its topological evolution in respect to the relevant parameter for all of the investigations mentioned-above. If the evolution exhibits the directional-veering phenomenon this type of the loss is classified as continuous otherwise it is assumed to be discontinuous. There exists also, reasonable grounds of this classification such as extremely different nucleation mechanisms underlying this two different transition types.

\section{Conclusion}\label{conclusion}
In this study, we have investigated the hysteretic behavior and dynamic nature of the critical phenomena which is observed for a site diluted Ising ferromagnet which has been defined with coordination number $z=3$ driven by an external oscillatory magnetic field by means of effective field theory based on a standard decoupling approximation and the time evolution of the system has been presented by utilizing a Glauber type stochastic process. Our starting point for the systematic review was to examine the frequency dispersion, coercivity and remanence coordinately for various field amplitude values at finite temperature below the Curie point of the static case on a pure Ising ferromagnet. In addition to the classification set forth in the literature, we categorized the frequency dispersion into four different types and labeled them according to the classification procedure which has been given in the text. We supported the relevant classification with hysteresis characteristics as well as the magnetization time series.

After that, in order to examine the effect of the dilution of site concentration on the dynamic nature of the system, firstly we presented the phase diagrams for the selected agents from three distinct types for various concentration values $c$ in the ($k_{B}T_{c}/J-\omega$) planes. In the immediate aftermath, the disorder induced evolution of the agents have been investigated at finite temperature corresponds along the route at $T\cong 1.045$ in the phase diagrams. According to our findings, a weak decreasing of the site concentration causes an increase in $\omega_{c}$ and vice versa. For a better view, we have depicted disorder dependency of the HLA, coercivity and the remanence for the aforementioned agents for various frequency values at $T\cong 1.045$. Also, our observations were the same, i.e., the relevant different type of dependencies has been observed similar to the frequency dependency classes.

EFT takes the standard MFA predictions one step forward by taking into account the single spin correlations which means that the thermal fluctuations are partially considered. Although all of the observations reported in this work shows that EFT can be successfully applied to such nonequilibrium systems in the presence of quenched site disorder, the true nature of the physical facts underlying the observations displayed in the system (especially the origin of the different distinct types of the parametric dependencies of the HLA related to the nucleation process and the HL residual area in the quasistatic limit) may be further understood with an improved version of the present EFT formalism which can be achieved by attempting to consider the multi spin correlations which originate when expanding the spin identities. We believe that this attempt could provide a treatment beyond the present approximation.

In conclusion, we hope that the results obtained in this work would shed light on the further investigations of the dynamic nature of the critical phenomena in disordered systems (e.g. scaling of the HLA by using the wide-range Lorentzian scaling function with four variabless in the effective-field case) and would be beneficial from both theoretical and experimental points of view.

\section*{Acknowledgments}
The numerical calculations reported in this paper were performed at T\"{U}B\.{I}TAK ULAKB\.{I}M (Turkish agency), High Performance and Grid Computing Center (TRUBA Resources) and this study has been completed at Dokuz Eyl\"{u}l University, Graduate School of Natural and Applied Sciences. One of the authors (B.O.A.) would like to thank the Turkish Educational Foundation (TEV) for full scholarship.

\newpage

\newpage

\newpage

\newpage

\newpage

\newpage

\newpage

\newpage


\section*{References}
\begin{thebibliography}{99}

\bibitem{steinmetz} C.P. Steinmetz, Trans. Am. Inst. Electr. Eng. 9 (1892) 3.
\bibitem{rao1} M. Rao, H. R. Krishnamurthy, R. Pandit, Phys. Rev. B 42 (1990) 856.
\bibitem{jung} P. Jung, G. Gray, R. Roy, P. Mandel, Phys. Rev. Lett. 65 (1990) 1873.
\bibitem{luse} C. N. Luse, A. Zangwill, Phys. Rev. E. 50 (1994) 224.
\bibitem{acharyya1} M. Acharyya, B.K. Chakrabarti, Physica A 192 (1993) 471.
\bibitem{acharyya2} M. Acharyya, B.K. Chakrabarti, R.B. Stincombe, J. Phys. A: Math. Gen. 27 (1994) 1533.
\bibitem{acharyya3} M. Acharyya, B.K. Chakrabarti, Phys. Rev. B 52 (1995) 6550.
\bibitem{fan} Z. Fan, Z. Jinxiu, Phys. Rev. Lett. 75 (1995) 2027.
\bibitem{chakrabarti} B. K. Chakrabarti, M. Acharrya, Rev. of Mod. Phys. 71 (1999) 847.
\bibitem{liu1} J.-M. Liu, H.L.W. Chan, C.L. Choy, C.K. Ong, Phys. Rev. B 65 (2002) 014416.
\bibitem{liu2} J.-M. Liu, H.L.W. Chan, C.L. Choy, Y.Y. Zhu, S.N. Zhu, Z.G. Liu, N.B. Ming, App. Phys. Lett. 79 (2001) 236.
\bibitem{suen} J.-S. Suen, J.L. Erskine, Phys. Rev. Lett. 78 (1997) 3567.
\bibitem{mayergoyz} I.D. Mayergoyz, Mathematical Model of Hysteresis, Springer-Verlag, Berlin, (1991).
\bibitem{binder} K. Binder, Phys. Rev. B 8 (1973) 3423.
\bibitem{gilmore} R. Gilmore, Phys. Rev. A 20 (1979) 2510.
\bibitem{agarwal} G.S. Agarwal, S.R. Shenoy ibid. 23 (1981) 2719.
\bibitem{acharyya4} M. Acharyya, Physica A 253 (1998) 199.
\bibitem{bruno} P. Bruno, G. Bayreuther, P. Beauvillain, C. Chappert, G. Lugert, D. Renard, J.P. Renard, J. Sneiden, J. Appl. Phys. 68 (1990) 5759.
\bibitem{tome} T. Tom\`{e}, M.J. de Oliveira, Phys. Rev. A 41 (1990) 4251.
\bibitem{glauber} R.J. Glauber, J. Math. Phys. 4 (1963) 294.
\bibitem{suzuki} M. Suzuki, R. Kubo, J. Phys. Soc. Jpn. 24 (1968) 51.
\bibitem{korniss} G. Korniss, P.A. Rikvold, M.A. Novotny, Phys. Rev. E 66 (2002) 056127.
\bibitem{berkoilako} G. Berkoilako, M. Grinfeld, Phys. Rev. E 76 (2007) 061110.
\bibitem{shi1} X. Shi, G. Wei, Phys. Lett. A 374 (2010) 1885.
\bibitem{idigoras} O. Idigoras, P. Vavassori, A. Berger, Physica B 407 (2012) 9.
\bibitem{jiang} Q. Jiang, H.N. Yang, G.C. Wang, Phys. Rev. B 52 (1995) 14911.
\bibitem{nowak} U. Nowak, J. Heimel, T. Kleinefeld, D. Weller, Phys. Rev. B 56 (1997) 8143.
\bibitem{robb} D.T. Robb, Y.H. Xu, O. Helling, J. McCord, A. Berger, M.A. Novotny, P.A. Rikvold, Phys. Rev. B 78 (2008) 134422.
\bibitem{kanuga} K. Kanuga, M. Cakmak, Polymer 48 (2007) 7176.
\bibitem{lo} W.S. Lo, R.A. Pelcovits, Phys. Rev. A 42 (1990) 7471.
\bibitem{acharyya5} M. Acharyya, Phys. Rev. E 58 (1998) 179.
\bibitem{sides} S.W., Sides, P.A., Rikvold, M.A., Novotny Phys. Rev. E 59 (1999) 2710.
\bibitem{zhu} H. Zhu, S. Dong, J.-M. Liu, Phys. Rev. B 70 (2004) 132403.
\bibitem{shi2} X. Shi, G. Wei, L. Li, Phys. Lett. A 372 (2008) 5922.
\bibitem{deviren} B. Deviren, O. Canko, M. Keskin, Chin. Phys. B 19 (2010) 050518.
\bibitem{punya} A. Punya, R. Yimnirun, P. Laoratanakul, Y. Laosiritaworn, Physica B 405 (2010) 3482.
\bibitem{lyuksyutov} I.F. Lyuksyutov, T. Nattermann, V. Pokrovsky, Phys. Rev. B 59 (1999) 4260.
\bibitem{nattermann1} T. Nattermann, V. Pokrovsky, V.M. Vinokur, Phys. Rev. Lett. 87 (2001) 197005.
\bibitem{nattermann2} T. Nattermann, V. Pokrovsky, Physica A 340 (2004) 625.
\bibitem{akinci} U. Akinci, Y. Yuksel, E. Vatansever, H. Polat, arXiV:1203.0639v1.
\bibitem{vatansever} E. Vatansever, B.O. Aktas, Y. Yuksel, U. Akinci, H. Polat, Jour. Stat. Phys. DOI:10.1007/s10955-012-0519-5.
\bibitem{honmura} R. Honmura, T. Kaneyoshi, J. Phys. C: Solid State Physics 12 (1979) 3979.
\bibitem{kaneyoshi1} T. Kaneyoshi, Acta Phys. Pol. A 83 (1993) 703.
\bibitem{tucker} J.W. Tucker, J. Magn. Magn. Mater. 102 (1991) 144.
\bibitem{zernike} F. Zernike, Physica A 7 (1940) 565.
\bibitem{balcerzak} T. Balcerzak, J. Magn. Magn. Mater. 97 (1991) 152.
\bibitem{kaneyoshi2} T. Kaneyoshi, R. Honmura, I. Tamura, E.F. Sarmento, Phys. Rev. B 29 (1984) 5121.
\bibitem{kaneyoshi3} T. Kaneyoshi, I. Tamura, E.F. Sarmento, Phys. Rev. B 28 (1983) 6491.

\bibitem{gunton} J.D. Gunton, M. San Miguel, P.S. Sahni, in Phase Transitions and Critical Phenomena, edited by C. Domb and J.L. Lebowitz (Academic, London, 1983), Vol. 8, p. 1.






\end{thebibliography}
\end{document}